\documentclass{svjour2}

\usepackage{amsfonts}
\usepackage{eucal}
\usepackage{amssymb}
\usepackage{amsbsy}
\usepackage[T1]{fontenc}
\newcommand{\scrif}{{\mathcal I}^{+}}
\newcommand{\scrip}{{\mathcal I}^{-}}
\newcommand{\T}{{\mathbb T}}

\newcommand{\scrT}{{\mathcal T}}
\newcommand{\const}{{\rm const}}
\newcommand{\scrS}{{\mathcal S}}
\newcommand{\geod}{{\boldsymbol{\gamma}}}
\newcommand{\Spin}{{\mathbb S}}
\newcommand{\omicron}{{o}}
\newcommand{\thorn}{\hbox{\TH}}
\newcommand{\rmi}{{\rm i}}
\newcommand{\rmd}{{\rm d}}

\journalname{General Relativity and Gravitation}
\begin{document}

\title{Angular momentum of isolated systems}

\author{Adam D. Helfer}

\institute{Department of Mathematics,
University of Missouri,
Columbia, MO 65211, U.S.A.\\
Tel.:  573-882-7283\\
Fax:  573-882-1869\\
E-Mail:  adam@math.missouri.edu }

\date{Received: date / Accepted: date}

\maketitle

\begin{abstract}
Penrose's twistorial approach to the definition of angular momentum at null
infinity is developed so that angular momenta at different cuts can be
meaningfully compared.  This is done by showing that the twistor spaces
associated with different cuts of $\scrif$ can be identified as manifolds (but
not as vector spaces). The result is a well-defined,
Bondi--Metzner--Sachs-invariant notion of angular momentum in a radiating
space--time; the difficulties and ambiguities previously encountered are
attached to attempts to express this in special-relativistic terms, and in
particular to attempts to identify a single Minkowski space of origins. Unlike
the special-relativistic case, the angular momentum cannot be represented by a
purely $j=1$ quantity $M_{ab}$, but has higher-$j$ contributions as well.
Applying standard kinematic prescriptions, these higher-$j$ contributions are
shown to correspond precisely to the shear.  Thus it appears that shear and
angular momentum should be regarded as different aspects of a single unified
concept.


\end{abstract}

\section{Introduction}

Energy--momentum and angular momentum (throughout this paper, `angular momentum'
will refer to the relativistic quantity) are of prime importance physically.
Practically, they are key conserved quantities in the analysis of specific
systems; theoretically, they are of fundamental significance, both classically
and in quantum physics.

Yet while these quantities are well-understood in special relativity, their
extension to general-relativistic situations is still in development. This is
because the {\em foundation} of the usual analysis --- the existence of the
Poincar\'e isometry group, generating the conserved quantities --- is absent:  a
general space--time will have no isometries.  We must expect, therefore, that in
generalizing momentum and angular momentum to curved space--time, we will be
forced to give up at least some of their familiar features, and perhaps to
reassess what their roots are.

For systems which are isolated in the sense of Bondi and Sachs, that is,
isolated in that one has a well-defined concept of what it means for radiation
to escape from them, the Bondi--Sachs energy--momentum is now accepted as
correct.  The success here has been closely connected with the existence of a
satisfactory four-dimensional group of asymptotic translations. For the angular
momentum of radiating systems, however, there have remained problems.  The
asymptotic symmetry group, the Bondi--Metzner--Sachs (BMS) group, (even though
it has the canonical four-dimensional translation subgroup) does {\em not} have
any physically preferred Poincar\'e subgroup, but rather an infinite-dimensional
family of them, apparently on equal footings. This issue often surfaces as an
ambiguity in defining a satisfactory space of origins with respect to which
angular momentum is to be measured.  One can in fact make a good argument that
it is {\em impossible } to identify a preferred physically natural space of such
origins.\footnote{Suppose the gravitational radiation is confined to several
finite intervals of retarded time. Then in the intervening, quiescent,
intervals, one has candidate spaces of origins, but there are in general gauge
mismatches (distortions by `supertranslations') between these quiescent periods
which prevent the identification of their different spaces of origins in any
natural and consistent way. Cf.~\cite{NP,GW,PR}.}

What properties should we look for in a definition of angular momentum for
isolated general-relativistic systems?  Certainly the definition itself should
be natural, which is to say it should respect the BMS symmetries. (It need not
be made in terms of the BMS generators, however.)  And of course it should be a
plausible extension of the special-relativistic definition (although perhaps
from an unorthodox perspective).  Finally, I think it is essential that angular
momenta at different cuts be unambiguously comparable, for that is what is
necessary for a concept of conservation, and a useful measure of exchange
between systems.

In this paper, I shall show that it is possible to achieve these goals by making
a definition of angular momentum directly on a twistor space.  The twistor
space, and the measure of angular momentum on it, have natural, invariant, and
unproblematic existences, which extend the definitions in special-relativistic
twistor theory.  (It is the attempts to pass from this twistor space to various
candidate `spaces of origins' which introduce difficulties and ambiguities.) 
But the  most important feature of the twistor space is that it is {\em
universal}, in that it depends only on those aspects of the asymptotic structure
which are common to all Bondi--Sachs systems.

The construction is a development of Penrose's quasilocal twistors applied at
future null infinity $\scrif$~\cite{P82,PR}.  Penrose showed that for each cut
$\scrS$ of $\scrif$, there exists a natural twistor space $\T (\scrS )$ with
familiar special-relativistic properties, most notably being a
four-complex-dimensional vector space. What I shall show here is that there is a
natural identification of all these twistor spaces as real manifolds.  This
means that we may say we have single twistor space $\scrT$ with many different
possible complex-linear structures, one for each cut.  The space $\scrT$ retains
enough structure that there is a meaningful twistorial definition of angular
momentum on it. As the origin-dependence of angular momentum becomes, in twistor
terms, a dependence on the choice of twistor, it is the {\em existence } of a
well-defined twistor space which is the solution of the problem.

While the twistor space $\scrT$ has no canonical linear structure, it does
canonically have a fibre-bundle structure, a bundle of affine complex planes
over a two-complex dimensional vector space.  This base space may be identified
with asymptotic spin space $\Spin _{A'}$. The elements of $\Spin _{A'}$ can be
used to specify the components of the angular momentum in question, and the
choice of point in the fibre corresponds to a choice of origin.

This means that if we think of measuring the angular momentum by specifying both
a component and an origin, then {\em fixing the component} one gets a sensible
definition (with a conventional origin-dependence); it is the behavior of this
construction as the component varies (in $\scrT$, as we pass from one fibre to
another) which is complicated.  We shall see that the components of the angular
momentum no longer constitute a pure $j=1$ quantity $M_{ab}$, but a more
complicated object with higher-$j$ contributions.

For instance, we will derive this formula for the spin:
\begin{equation}\label{spinin}
 {\rm spin} (\hat{\bf r}) =s_{\rm v}^a{\hat{\bf r}}_a
 +M\Im\lambda ({\hat{\bf r}})\, ,
\end{equation}
where $\hat{\bf r}$ is a unit spatial vector representing the direction in which
the spin is to be measured, the vectorial part of the spin is $s_{\rm v}^a$, the
Bondi--Sachs mass is $M$, and $\Im\lambda$ is an angular potential for the
magnetic part of the Bondi shear, with components for $j\geq 2$.  In other
words, the spin can be measured in different directions, but these measurements
do not `integrate up' to give simply a spin-vector; they give a more complicated
angular dependence.  Note that here (as throughout this paper) the quantity $j$
refers, not to the values that angular momentum takes, but to the functional
dependence of the angular momentum on direction.

The formula~(\ref{spinin}) is interesting for a number of reasons. First, it
gives an explicit and transparent role to the magnetic part of the shear.
(Investigations have repeatedly turned up a sensitivity of angular-momentum
constructions to the magnetic shear, but the precise role has been difficult to
pin down.)  Second, the remarkable simplicity of this formula can fairly be
taken as evidence in favor of the present approach. Third, this result is
intuitively plausible, because in special relativity it is well-known that
$M^{-1}s_{\rm v}^a$ can be interpreted as a displacement of the center of mass
into the complex and in twistor theory $\Im\lambda$ has a similar interpretation
as a displacement of the cut into the complex.

One can also apply standard kinematic formulas to identify the center of mass,
and it has a non-vectorial part which may be identified  with the angular
potential $\Re\lambda$ for the electric part of the Bondi shear. Thus it is the
Bondi shear which controls the non-Poincar\'e character of the angular momentum,
the electric and magnetic parts of the shear contributing non-Poincar\'e
behavior to the center-of-mass and spin, respectively.

These results suggest that the shear forms a sort of angular momentum,
essentially the $j\geq 2$ parts of the angular momentum, and that, in passing
from special to general relativity, the mixing of gravitational radiation and
matter mixes `ordinary' ($j=1$) angular momentum and radiative modes of the
field, so that the `correct' understanding of angular momentum embraces the two
concepts.

The most important consequence of the close identification of angular momentum
with shear is that gravitational radiation may carry off angular momentum
(including $j=1$ parts) as a {\em first-order} effect (in the Bondi news),
whereas energy--momentum is radiated as a second-order effect. Thus the
radiation of angular momentum may be a more important effect for, and a more
significant constraint on, many systems than is the radiation of
energy--momentum.

Still, the effects are not generally expected to be large except in highly
relativistic circumstances (or perhaps cumulatively over long times or large
volumes).  They are typically characterized by the length scale set by the Bondi
shear --- more precisely, by the magnetic part of the shear, and by changes in
the electric part of the shear.  This scale is typically of order $\sim
(GM/c^2)(v/c)^n$, where $M$ is a characteristic mass and $v$ a characteristic
velocity and $n\geq 2$.\footnote{This is of course a very crude statement and is
only useful at the coarsest level.  Given any specific system, one needs to
think about what the appropriate choices are for $M$ and $v$. For example, in
cases where there is a change in electric shear the correct characteristic
velocity $v$ may be something like the square root of the change in the squared
velocity of a component of the system.} Even for relativistic velocities, this
is only of order the Schwarzschild radius of the mass.  Thus while it is to be
hoped that eventually astrophysical measurements will be  refined enough to
warrant the use of general-relativistically-corrected angular momenta, the main
interest at present is theoretical.

One would also expect the ideas here to give quantum-gravitational corrections
to angular momentum and spin.  These effects do not appear to be significant for
individual microscopic systems, since the corrections are typically of the order
$(E/E_{\rm Pl})^2\hbar$, where $E$ is the energy of the system and $E_{\rm Pl}$
is the Planck energy.  However, the development of a quantum theory of spin
incorporating the ideas here could very well have implications for the final
stages of black hole evaporation, because it could constrain the possible
transitions.

\subsection{Real twistors, magnetic shear and origins}

This subsection deals with the reality structure on twistor space. While this
will figure essentially in the paper, the reader not specially interested in
twistor theory can safely skip this part, referring back to it later if
necessary.

While the ideas sketched above present an accurate outline of the main points in
conventional space--time terms, they do leave out one crucial element of the
twistor structure, its reality properties.  It turns out that these are closely
tied to the magnetic part of the shear, to the failure of a model Minkowski
space of origins to exist, and to differences between the structures of $\scrT$
and $\T (\scrS )$.

In order to explain this, a quick outline of the special-relativistic case is in
order.  There, twistor space has a Hermitian form of signature $+{}+{}-{}-$, and
the twistors $Z^\alpha$ which are null ($Z^\alpha {\overline Z}_\alpha =0$) with
respect to it are called `real'.  The points in Minkowski space may be
identified with the totally real (that is, totally null) complex two-planes, and
thus real twistors are involved centrally in definition of origins for angular
momentum.

In the general-relativistic case, we shall find that there is a natural
definition of a quantity $\Phi$ on $\scrT$ analogous to the form $Z^\alpha
{\overline Z}_\alpha$ on special-relativistic twistor space, and one can define
real twistors with respect to $\Phi$ and thus develop a theory of angular
momentum and spin. Of course, since $\scrT$ has no preferred linear structure,
one cannot properly say that $\Phi$ is a (squared) norm.  However, the failure
is not just a semantic nicety.

Given any cut $\scrS$, there is a natural identification of the twistor space
$\scrT$ with Penrose's $\T (\scrS )$, which could then be used to give $\scrT$
an $\scrS$-dependent complex-linear structure.  However, in general the function
$\Phi$ does {\em not} respect that linear structure --- that is, in general, the
function $\Phi$ is {\em not} the restriction of any Hermitian form on $\T (\scrS
)$, but is more strongly nonlinear. This means that in general one {\em cannot}
find totally $\Phi$-real two-planes in $\T (\scrS )$, and thus one cannot
reconstruct from $\T (\scrS )$ a Minkowski space of origins. In fact, we shall
see that the condition for $\Phi$ to arise from a Hermitian form on $\T (\scrS
)$ is {\em precisely} that $\scrS$ should have no magnetic shear.  Thus it is
precisely the presence of magnetic shear on $\scrS$ which prevents the
identification of a Minkowski space of origins from $\scrT$.

\subsection{Organization}

The plan of the paper is this.  The next section reviews twistorial kinematics
and Penrose's construction.   Section~\ref{ident} is the heart of the paper; it
explains how the twistor spaces at different cuts may be identified, and what
the structure of the resulting nonlinear twistor space is. Section~\ref{kinte}
explains how this space may be used to compute angular momentum. There the
formulas for the spin and center of mass are deduced, as well as a flux law.
Section~\ref{lint} explains how the present (non-Poincar\'e) angular momentum is
connected to the familiar (Poincar\'e) special-relativistic one, by discussing
the case of linearized gravity.  An account of the number of constants of motion
produced by this framework, and the relation of the relation of angular momentum
to gravitational degrees of freedom, is given in section~\ref{constm}.  The
final section recapitulates the paper's main conclusions.

{\em Notation, conventions and background.  } The notation and conventions are
those of Penrose and Rindler~\cite{PR}, in which also all concepts not explained
here can be found.  This paper assumes a familiarity with two-spinors and
spin-coefficients, as well as Penrose's conformal treatment of $\scrif$. All
computations at $\scrif$ are in terms of the conformally rescaled quantities. 
Factors of $G$ are given explicitly, but some factors of $c$ are suppressed.

In this paper, we work at future null infinity $\scrif$.  However, parallel
results, applying to incoming radiation, could be obtained at past null infinity
$\scrip$, by reversing the sense of time.

For a review of the problems of defining angular momentum and related issues,
see the article by Szabados~\cite{S}. Another sort of curved twistor space was
introduced by Penrose~\cite{P76}; this was essentially equivalent to the
${\mathcal H}$-space of Newman~\cite{KLNT}, which had roots in work on angular
momentum as well.

\section{Twistorial kinematics and Penrose's construction}

This section summarizes the main points of twistor theory which will be used.
Section~\ref{tms} can be safely skipped by those familiar with twistor theory.
The same is true for most of section~\ref{tcs}, although the final two
paragraphs do contain some new comments about the case of cuts with magnetic
shear.  Section~\ref{kt} should be read, because the material is presented from
a slightly unconventional vantage adapted to later arguments.

\subsection{Twistors in Minkowski space}\label{tms}

In Minkowski space, twistors can be regarded as solutions of the {\em
twistor equation}
\begin{equation}
\nabla ^{A'(A}\omega^{B)}=0\, .
\end{equation}
It can be shown that there is a four-complex-dimensional family $\T$ of
solutions, each of the form $\omega ^A(x)={}_0\omega ^A-\rmi x^{AA'}\pi _{A'}$,
where ${}_0\omega ^A$ and $\pi _{A'}$ are fixed spinors.  A twistor, when it is
thought of as an element of this space (that is, when its character as a spinor
field is not emphasized) is generally denoted by $Z^\alpha$, and one sometimes
writes $Z^\alpha =({}_0\omega ^A,\pi _{A'})$.  Under Lorentz motions, the
spinors transform conventionally.  Under a translation $x^a\mapsto x^a+k^a$, we
have $\omega ^A(x+k)={}_0\omega ^A -\rmi (x^{AA'}+k^{AA'})\pi _{A'}$, so
$({}_0\omega ^A,\pi _{A'})\mapsto ({}_0\omega ^A-\rmi k^{AA'}\pi _{A'},\pi
_{A'})$.  This means that the projection $Z^\alpha\mapsto \pi _{A'}$ is
Poincar\'e-covariant.  Complementarily, the twistors with $\pi _{A'}=0$ have
${}_0\omega ^A$ translation-invariant and are said to lie at infinity.

The following twistorial structures are also important.  There is an alternating
twistor $\epsilon _{\alpha\beta\gamma\delta}$ and a norm $Z^\alpha{\overline
Z}_\alpha ={}_0\omega ^A{\overline\pi}_A +\overline{{}_0\omega}^{A'}\pi _{A'}$. 
(These are both conformally invariant.) A twistor with $Z^\alpha{\overline
Z}_\alpha =0$ is said to be {\em null} or {\em real}.  Additionally, there is a
(Poincar\'e-invariant) {\em infinity twistor}
\begin{equation}
 I^{\alpha\beta}=\left[\matrix{\epsilon ^{AB}&0\cr 0&0\cr}\right]\,
.
\end{equation}
For any skew bitwistor $X^{\alpha\beta}=-X^{\beta\alpha}$, its dual is defined
by $X_{\alpha\beta}=(1/2)\allowbreak\epsilon
_{\alpha\beta\gamma\delta}\allowbreak X^{\gamma\delta}$.  A skew bitwistor is
real iff $X_{\alpha\beta}={\overline X}_{\alpha\beta}$.  In particular, the
infinity twistor is real.

Any point $x$ in Minkowski space determines a complex two-plane of twistors,
those for which $\omega ^A(x)=0$, and all twistors on this plane are null.
Conversely, any complex two-plane in $\T$ all of whose elements are null
determines a point in Minkowski space (or a point at infinity).  The points at
infinity in Minkowski space are precisely those whose twistor planes include at
least one twistor at infinity. Correspondingly, a null twistor vanishes exactly
on a null geodesic in Minkowski space, along which its spinor $\pi _{A'}$ is
tangent (or the twistor corresponds to a limit at infinity of this situation).

\subsection{Kinematics twistorially}\label{kt}

If $P_a$ and $M_{ab}=\mu _{A'B'}\epsilon _{AB}+{\overline\mu}_{AB}\epsilon
_{A'B'}$ are the momentum and angular momentum of a special-relativistic system,
they define a {\em kinematic twistor}, whose spinor components are given by
\begin{equation}\label{ktwc}
 A_{\alpha\beta} =\left[\matrix{0&P_A{}^{B'}\cr P^{A'}{}_B&2\rmi\mu
^{A'B'}\cr}\right]\, .
\end{equation}
That is, the angular momentum has exactly the correct origin-dependence for
$A_{\alpha\beta}$ to transform as a (dual bi-)twistor.

For any twistor $Z^\alpha$, then, the combination
\begin{equation}\label{faeq}
 A(Z)=A_{\alpha\beta} Z^\alpha Z^\beta =2\rmi\mu ^{A'B'}\pi _{A'}\pi
_{B'} +2P_A{}^{A'}\omega ^A\pi _{A'}
\end{equation}
is invariant, that is, the origin-dependences of $\mu ^{A'B'}$ and $\omega ^A$
cancel out.  In fact, this quantity is simply $2\rmi$ times the  $\pi
_{A'}$-component (or, more properly, the $\pi _{A'}\pi _{B'}$-component) of the
spinor form of the relativistic angular momentum at any point $x$ where $\omega
^A(x)=0$. We may thus regard $A(Z)$ as measuring the angular momentum, by
choosing $Z^\alpha =(\rmi x^{AA'}\pi _{A'},\pi _{A'})$, where $x$ is the point
at which to evaluate the angular momentum and $\pi _{A'}$ determines the
component in question.  (By polarization, or equivalently, by differentiating
with respect to $\pi _{A'}$, one can get all components of the angular momentum
this way.)  Thus a knowledge of $A(Z)$ on null twistors is equivalent to a
knowledge of angular momentum, the choice of twistor coding both the origin with
respect to which one is measuring and the component measured.

We shall have also to consider the polarized form $A(Z,\acute
Z)=A_{\alpha\beta}Z^\alpha {\acute Z}^\beta$.  For us, it is most natural to
think of this as a sort of two-point function on twistor space.  It is possible,
of course, to interpret this in more conventional special-relativistic terms; a
brief calculation shows that
\begin{equation}\label{polar}
 A(Z,\acute Z )=2\rmi \mu ^{A'B'}(x_{\rm
av})\pi
_{A'}{\acute\pi}_{B'} -(\rmi /2)P_ax_{\rm diff}^a\pi
_{B'}{\acute\pi}^{B'}\, ,
\end{equation}
where $x^a_{\rm av}=(1/2) (x^a+{\acute x}^a)$ and $x_{\rm diff}^a= x^a-{\acute
x}^a$. In this formula, the points $x^a$, ${\acute x}^a$ are arbitrary points on
the twistors $Z$, $\acute Z$. They are thus subject to the freedoms $x^a\to
x^a+\xi{\overline\pi}^A\pi ^{A'}$, ${\acute x}^a\to{\acute x}^a+\acute\xi
{\acute{\overline\pi}}^A{\acute\pi}^{A'}$.  There is a one-parameter family of
these for which the factor $P_ax^a_{\rm diff}=0$, and it defines a timelike
world-line of values of $x^a_{\rm av}$ with tangent $P^a$. Thus one can
interpret $A(Z,\acute Z)$ as a component of the angular momentum at any point on
this world-line.  However, this interpretation will not be available in the
generic general-relativistic setting, because we will not really have a model
Minkowski space in which to envision world-lines.

We are led to think of the angular momentum as actually being the functions
$A(Z)$, $A(Z,\acute Z)$ on twistor space.  Thus we think of evaluating the
angular momentum, not by taking particular components at a particular
space--time point, but by evaluating $A(Z)$, $A(Z,\acute Z)$ on twistor space
(with a distinguished role played by the values at twistors with $Z^\alpha
{\overline Z}_\alpha =0$). This will be our point of view in what follows.  This
twistorial description of angular momentum remains valid even when it is
difficult to define a space of origins. Since each null twistor corresponds to a
null geodesic with tangent spinor,  the function $A(Z)$ is just the component of
the angular momentum determined by $\pi _{A'}$ evaluated along the geodesic. The
function $A(Z,\acute Z)$ can be also be interpreted as a sum of the angular
momentum at the average of any two points on the geodesics plus a correction due
to the difference between the two points (cf. the previous paragraph), but it is
really most natural to think of it as simply a two-point function on the space
of real twistors.

Twistor expressions for the spin are a little involved.  There are several ways
of doing it; we give one which will be well-suited to later generalization. 
Suppose that $Z^\alpha$, ${\acute Z}^\alpha$ are both real and also satisfy
$A_{\alpha\beta}Z^\alpha Z^\beta =0$, $A_{\alpha\beta}{\acute Z}^\alpha {\acute
Z}^\beta =0$.  Then by direct calculation, one can verify the following
identity:
\begin{eqnarray}\label{spin}
\Re A_{\alpha\beta} Z^\alpha {\acute Z}^\beta /I_{\gamma\delta}Z^\gamma
{\acute Z}^\delta = (2P^{AA'}{\overline\pi}_A\pi _{A'}
P^{BB'}{\acute{\overline\pi}}_{B}{\acute\pi}_{B'})^{-1}\cdot\nonumber\\
\left(
 S^{AA'}{\overline\pi}_A\pi
_{A'}P^{BB'}{\acute{\overline\pi}}_B{\acute\pi}_{B'} -
S^{AA'}{\acute{\overline\pi}}_A{\acute\pi}_{A'}
P^{BB'}{\overline\pi}_B\pi _{B'}\right)\, ,
\end{eqnarray}
where $S_a=(1/2)\epsilon _{abcd}P^bM^{cd}$ is the Pauli--Luba\'nski spin
covector.  This identity allows one to recover the spin in terms of real
twistors.  (It is not quite obvious that given $\pi _{A'}$ and
${\acute\pi}_{A'}$ there do exist real twistors $Z^\alpha$, ${\acute Z}^\alpha$
with $A_{\alpha\beta}Z^\alpha Z^\beta =A_{\alpha\beta}{\acute Z}^\alpha {\acute
Z}^\beta =0$, but this can be deduced from~(\ref{faeq}).)

\subsection{Twistors on cuts of $\scrif$}\label{tcs}

In a curved space--time, the twistor equation is in general over-determined and
has only the trivial solution.  However, in favorable circumstances one can
obtain twistor spaces by considering just certain components of the equation. If
$\scrS$ is a cut of $\scrif$, the components of the twistor equation involving
only derivatives tangent to $\scrif$ are
\begin{equation}
\eth ' \omega ^0=0\, ,\qquad\eth\omega ^1=\sigma \omega ^0\, ,
\end{equation}
where we use the standard spin-coefficient formalism for the geometry expressed
in conformally rescaled terms (cf.~\cite{PR}). This system of equations defines a
four-complex-dimensional vector space $\T (\scrS )$, which turns out to have an
infinity twistor and alternating twistor obeying the same algebra as those in
Minkowski space.

The kinematic twistor, deduced by correspondence with linear theory, is
\begin{equation}
A(Z)=A_{\alpha\beta}Z^\alpha Z^\beta ={{-\rmi}\over{4\pi G}}\oint
 \left\{ \psi _1 (\omega ^0)^2 +2(\psi _2 +\sigma \dot{\overline\sigma}
)\omega
^0\omega ^1\right\} \, \rmd
{\scrS}\, ,
\end{equation}
where the overdot represents differentiation with respect to $u$ and an
integration by parts has been used to simplify the expression somewhat.

To complete the treatment of kinematics, one also needs a norm (and to verify
that the other quantities have the appropriate reality properties with respect
to it).  The natural thing to do is to express $Z^\alpha {\overline Z}_\alpha$
in terms of the spinor fields $\omega ^0$, $\omega ^1$.  This can be done, but
one finds that the resulting expression, while constant in Minkowski space, or
on a cut with purely electric shear, is not constant if the shear has a magnetic
component.  (Recall that the shear $\sigma$, being a spin-weight two quantity,
can always be expressed as $\sigma =\eth ^2\lambda$ for an ordinary function
$\lambda$.  The electric and magnetic parts of the shear are $\sigma _{\rm
el}=\eth ^2\Re\lambda$ and $\sigma _{\rm mag}=\eth ^2\rmi\Im\lambda$.) Penrose
suggested, with some reservations, averaging over the cut with respect to the
metric defined by the Bondi--Sachs energy--momentum.  Then one does get the
proper special-relativistic twistor algebra.

There are two concerns about this norm (in the cases where there is a magnetic
component to the shear) which should be raised here.  The first is its strong
cut-dependence, which makes it hard to  understand how to compare angular
momenta at different retarded times.  The second is that, were one to adopt it,
one would have to regard the physical space--time as displaced into the complex
relative to the reality structure on the twistor space~\cite{PR}, and this would
mean (for example) that a massless test particle on a light ray, which by local
measurements had zero helicity, would have to be ascribed a non-zero helicity
according to the twistorial definition.

\section{Identifying the twistor spaces}\label{ident}

I show here how the twistor spaces at the different cuts of $\scrif$ may be
identified.  While the ideas are not difficult, it is perhaps well to begin by
cautioning against a possible preconception. On each cut the twistors are
defined as spinor fields satisfying certain equations, so one might be tempted 
to think that the twistors could be realized as spinor fields over all of
$\scrif$, which simply restrict to the correct values on any cut.  However, the
twistors have a more delocalized existence, taking well-defined values as spinor
fields only when an entire cut (and not simply a point on the cut) is specified.

\subsection{The field $\omega ^0$}

The components $\omega ^0$, $\omega ^1$, and the two components of the
two-surface twistor equation, are not on the same footing as far as the
delocalization just discussed goes. It turns out that the space of $\omega ^0$s
satisfying $\eth '\omega ^0=0$ on any cut {\em does} have a well-defined
existence as a field on all of $\scrif$.  This feature is a well-known part of
Penrose's construction, the space of such $\omega ^0$ being identified with
$\Spin _{A'}$, the space of primed dual twistors `at infinity', and the
projection $(\omega ^0,\omega ^1)\mapsto \omega ^0$ providing a fibration $\T
(\scrS )\to\Spin _{A'}$ of the twistor space on a given cut to this space. The
$\omega ^0$ spinors in this case satisfy $\thorn '\omega ^0=0$ (as well as $\eth
'\omega ^0=0$), which can be thought of as providing a propagation equation from
one cut to another.  These two equations may be expressed jointly as $\iota
_{A}\iota _B\nabla ^{AA'}\omega ^B=0$, which makes the cut-independence of the
system clear, since $\iota ^A$ is (apart from scaling) independent of the Bondi
system.

There will be a consequence of this which will figure centrally in what
follows.  This is that for any admissible $\omega ^0$ field which is not
identically zero, there is a unique generator $\gamma (\omega ^0)$ of $\scrif$
on which $\omega ^0$ vanishes.  (This is a well-known consequence of the
equation $\eth '\omega ^0=0$, given that $\omega ^0$ has spin-weight $-1/2$,
together with the equation $\thorn '\omega ^0=0$, which transports $\omega ^0$
up the generators.)  The generator on which $\omega ^0$ vanishes is also
independent of the choice of Bondi system, as can be seen by writing the
vanishing equation in the form $\omega ^A\iota _A=0$.

\subsection{The field $\omega ^1$}

We have seen that, for each twistor, the field $\omega ^0$ has a well-defined
cut-independent existence.  However, the situation for $\omega ^1$ is more
complicated.

In order to  treat twistors without having in each instance to take up their
global behavior over a cut, we take advantage of the fact that each twistor can
be specified, at any point of the cut, by suitable data.  These data are most
commonly taken to be the values $(\omega ^0,\omega ^1,\pi _{0'},\pi _{1'})$,
where $\pi _{0'}=\rmi\eth '\omega ^1-\rmi \rho\omega ^0$, $\pi
_{1'}=\rmi\eth\omega ^0$.\footnote{For convenience, we choose $\rho '=0$.} We
may thus, in order to ask how to transport a twistor from one cut to another,
simply ask how to identify its data at one point on the first cut with data at
another point on the second cut.

Our prescription will essentially consist of transporting the twistor along the
generator $\gamma (\omega ^0)$ on which $\omega ^0$ vanishes; this generator
must intersect each cut in a single point. (The case of identically vanishing
$\omega ^0$ will be seen later to be determined by continuity, and also to agree
with Penrose's definition of the spin space $\Spin ^A$.)

In order to determine the correct transport equations, we recall the
formula for the {\em local twistor connection}:
\begin{equation}\label{ltc}
D_{BB'}\left(\omega ^A ,\pi _{A'}\right) =\left( \nabla _{BB'}\omega ^A
+\rmi\epsilon _B{}^A\pi _{B'},\nabla _{BB'}\pi _{A'}+\rmi P_{ABA'B'}\omega
^A\right)\, ,
\end{equation}
where $P_{ab}=\Phi _{ab}-\Lambda g_{ab}=(1/12)Rg_{ab}-(1/2)R_{ab}$. The equation
$D_{BB'}\left(\omega ^A ,\pi _{A'}\right)\allowbreak =0$ has, in a conformally
flat space--time, locally a four-complex-dimensional family of solutions which
comprise twistor space.  At $\scrif$ when gravitational radiation is present,
this equation is over-determined and does not have a full four-dimensional
family of solutions.  However, we shall {\em only} use components where the
derivative is tangent to a generator of $\scrif$ and we shall {\em only} enforce
it on the single generator $\gamma (\omega ^0)$.  Thus we have a system of
ordinary differential equations on a line, and there are no integrability
restrictions.

The components of (\ref{ltc}) in question are
\begin{equation}\label{invtrnsp}
 n^bD_b\left(\omega ^A,\pi _{A'}\right) =0\mbox{ on }\gamma
(\omega ^0)\, .
\end{equation}
In this form, they are manifestly independent of the choice of Bondi
system.  In spin-coefficient notation, they become
\begin{equation}\label{trnsp}
 \thorn '\omega ^0=0\, ,\quad \thorn '\omega ^1=\eth\omega ^0\,
,\quad
 \thorn '\eth '\omega ^1=0\, ,\quad \thorn '\eth\omega ^0=0\mbox{ on }
\gamma (\omega ^0)\, .
\end{equation}
We shall use these now to derive an explicit transformation for twistors from
one cut to another.  For simplicity, we take these here to be two $u=\const$
cuts of the same Bondi system, although this is not necessary.

The twistors on the cuts may be conveniently given.  We let $\lambda$ be any
function satisfying $\eth ^2\lambda =\sigma$; then the $\omega ^1$ fields on the
cut have the form
\begin{equation}
\omega ^1=\omega ^0\eth\lambda -\lambda \eth\omega ^0 +\xi\, ,
\end{equation}
where $\xi$ is any solution to $\eth\xi =0$.  (This observation is due to
K. P. Tod.)  The space of such $\xi$ is spanned by $\eth\omega ^0$
and ${\overline\omega}^{0'}$, and so
\begin{equation}\label{twfrm}
\omega ^1=\omega ^0\eth\lambda -\lambda \eth\omega ^0
+\alpha\eth\omega ^0+\beta{\overline\omega}^{0'}\, ,
\end{equation}
where $\alpha =\alpha (u)$ and $\beta =\beta (u)$ are to be determined.
Substituting (\ref{twfrm}) into (\ref{trnsp}) and integrating, we find
\begin{eqnarray}
 \alpha &=& u-u_0 +\lambda (u,\gamma )-\lambda (u_0,\gamma) +\alpha
_0\label{trnsn}\\
 \beta &=& 
  {{\eth\omega ^0(\gamma
)}\over{\eth'{\overline\omega}^{0'}(\gamma )}}
  \left(\eth '\lambda
(u,\gamma )-\eth '\lambda (u_0,\gamma )\right) +\beta _0\, ,\label{trnsnb}
\end{eqnarray}
where $\alpha (u_0)=\alpha _0$, $\beta (u_0)=\beta _0$, and we have abused
notation slightly by writing the generator $\gamma$ as the argument of functions
(rather than the angular variables determining $\gamma$).

These formulas allow us to identify the twistor spaces $\T (\scrS )$ at
different cuts, so we may say that we have a single twistor space $\scrT$.  For
any fixed $u_0$, we may view $\alpha (u_0)$ and $\beta (u_0)$ (together with a
coordinatization of the $\omega ^0$ fields)  as providing coordinates on $\scrT$
(and identifying it with $\T (\scrS _0)$, where $\scrS _0$ corresponds to
$u_0$).  As $u$ varies, the transport formulas give us transition functions to
other coordinate systems $(\alpha (u),\beta (u))$. The formulas are generally
nonlinear, because of the nonlinear dependence of the terms on $\gamma$, the
generator on which $\omega ^0$ vanishes.

So far we have excluded the case of twistors with identically vanishing $\omega
^0$.  To obtain coordinate charts including such twistors we choose some
coordinatization of the space of $\omega^0$ fields and also of the $\xi$ fields;
together these coordinatize the twistors according to~(\ref{twfrm}).  Whatever
coordinatization of these we choose, we will have
\begin{eqnarray}
  \xi (u)-\xi (u_0) =\left( u-u_0+\lambda(u,\gamma )-\lambda
(u_0,\gamma)\right)\eth\omega ^0 +\nonumber\\ \label{salmon}
  {{\eth\omega ^0(\gamma
)}\over{\eth'{\overline\omega}^{0'}(\gamma )}}
  \left(\eth '\lambda
(u,\gamma )-\eth '\lambda (u_0,\gamma )\right){\overline\omega}^{0'}\, ,
\end{eqnarray}
by (\ref{twfrm},\ref{trnsn},\ref{trnsnb}),  and this will determine the
transition functions.  Since this has a well-defined limit (zero) as $\omega ^0$
is taken identically to zero, the transition functions extend to be $C^0$ there.

\subsection{Structure of $\scrT$}

We have just seen that $\scrT$ is a smooth manifold except where $\omega ^0$
vanishes identically, where it is of class $C^0$. The transition functions
respect the fibration over $\Spin _{A'}$, so the space $\scrT$ fibres over
$\Spin _{A'}$.  Each fibre has the structure of an affine complex two-plane, for
the transition functions only act by translations.  Each choice of cut provides
a trivialization of the bundle, identifying $\scrT$ with $\T (\scrS )$. There is
a well-defined scaling action $(\omega ^0,\omega ^1,\pi _{0'},\pi _{1'})\mapsto
k(\omega ^0,\omega ^1,\pi _{0'},\pi _{1'})$.

We shall take as the infinity twistor $I(Z,\acute Z)={\acute\omega}^0\eth\omega
^0-\omega ^0\eth{\acute\omega}^0$ (which is constant over $\scrif$), in
agreement with Penrose. There is a well-defined real function on $\scrT$ which
naturally extends the definition of the (squared) norm, that is the quantity
$\Phi =\omega ^A{\overline\pi}_A+\pi _{A'}{\overline\omega }^{A'}$ evaluated at
$\gamma (\omega ^0)$ (and equal to zero if $\omega ^0$ vanishes identically).
Even on a specific cut $\scrS$, where $\T (\scrS )$ has a complex-linear
structure, the form $\Phi$ is {\em not} in general Hermitian, because the
expression $\omega ^A{\overline\pi}_A +\pi _{A'}{\overline\omega}^{A'}$ must be
evaluated at the point $\gamma (\omega ^0)$ depending on the twistor in
question; this introduces a significant additional nonlinearity.  In fact (as
Penrose noted), the condition that this expression be constant over $\scrS$ is
precisely that $\scrS$ be purely electric.  Our function $\Phi$ thus reduces to
a Hermitian one (and agrees with Penrose's) on a cut precisely if the cut is
purely electric; otherwise it is a more complicated object.

This strongly nonlinear $\Phi$ has two  advantages.  First, it is independent of
the choice of cut, and this means that we have a universal notion of real
twistors, and thus a way of comparing angular momenta on different cuts of
$\scrif$.  Second, it provides a direct link to space--time geometry, for with
this definition of $\Phi$ we may identify the real twistors with the null
geodesics (together with their parallel-propagated tangent spinors) meeting
$\scrif$, in the usual way.  The function $\Phi$ does restrict to a Hermitian
form on those twistors with the field $\omega ^0$ fixed up to proportionality,
because in this case one does not need to vary $\gamma (\omega ^0)$. If we fix
also the scale of $\omega ^0$ (and require $\omega ^0$ not identically zero),
then the set of all null twistors with this $\omega ^0$ field is evidently a
real three-dimensional affine space.  In the special-relativistic case, this
space would be identified with Minkowski space modulo the translations by
${\overline\pi}^A\pi ^{A'}$, where $\pi _{A'}$ stands for $\omega ^0$ regarded
as an element of $\Spin _{A'}$.  We thus get, for each $\pi _{A'}\not=0$, a
space which can be regarded as a space of origins modulo translations by
multiples of ${\overline\pi}^A\pi ^{A'}$. As we expect the component of the
angular momentum in the direction $\pi _{A'}$ to be independent of such
translations, we will have, for this component, as much of a well-defined space
of origins as is necessary for the definition of angular momentum.

\section{Kinematics on $\scrT$}\label{kinte}

As our construction allows us to identify $\scrT$ with Penrose's $\T (\scrS )$
on a cut $\scrS$, we may take over Penrose's definition of the kinematic twistor
directly to $\scrT$,
\begin{equation}\label{kint}
A_\scrS (Z)={{-\rmi}\over{4\pi G}}\oint\left\{ \psi _1(\omega ^0)^2
 +2(\psi _2+\sigma\dot{\overline\sigma} )\omega ^0\omega ^1\right\}\,
\rmd \scrS\, .
\end{equation}
This defines the `unpolarized' form, corresponding to the Minkowski quantity
$A_{\alpha\beta}Z^\alpha Z^\beta$. A polarized form, corresponding to
$A_{\alpha\beta}Z^\alpha{\acute Z}^\beta$, can also be defined.  While these
forms are quadratic and bilinear, respectively, as functions on $\T (\scrS )$,
one cannot say they have these properties as functions on $\scrT$, because of
the lack of a linear structure on that space.

Even restricting our attention to one cut, however, the linear structure on $\T
(\scrS )$ is in generic circumstances much less significant than it is in
special relativity, for the special-relativistic formalism for angular momentum
is recovered only when the shear is purely electric. If the shear is not purely
electric, one cannot recover a suitable Minkowski space of origins with respect
to which one might define angular momentum in familiar Poincar\'e terms.  We are
thus forced to confront, even on a single cut, a non-Poincar\'e character of the
angular momentum:  what does $A_\scrS (Z)$ mean?

Briefly, we may interpret $A_\scrS (Z)$, for real $Z$, as the $\pi
_{A'}$-component of the angular momentum about the geodesic defined by $Z$ (or
really about the point where $Z$ meets $\scrif$, since the construction is
really defined at $\scrif$). This interpretation is valid whether $Z$ meets
$\scrif$ in $\scrS$ or not, and whether $Z$, if it does meet $\scrS$, is a
member of the congruence of null geodesics meeting $\scrS$ orthogonally. A
fuller understanding will be developed by investigating the spin and center of
mass.

\subsection{Magnetic shears and spin}\label{magsec}

As noted above, when a magnetic component of the shear is present, the
definition of the norm adopted here in general precludes the identification of a
Minkowski space of origins.  It is natural to ask what happens to the spin in
this situation, for in special relativity the spin is origin-independent.  Thus
we may hope to get an origin-independent spin in general relativity. It is not
at all obvious, however, that it is possible to do so.  The most common
special-relativistic approaches to spin have no obvious analogs here, because
these approaches begin from the angular momentum evaluated at an origin. 
Remarkably, however, there is a suitable such definition. We saw in
section~\ref{kt} that in special relativity there is an expression for the spin
in twistorial terms, and it turns out that the same expression results in an
origin-independent definition of spin here.  The definition is more complicated
than in special relativity, though, in that the spin is not (if magnetic shear
is present) purely vectorial ($j=1$), but has higher-$j$ components as well.

In order to derive it, let us begin by working out $A_\scrS (Z,\acute Z )$ in
some detail.  For simplicity we take the time axis of the Bondi system aligned
with the Bondi--Sachs energy--momentum, and assume that $\lambda$ has no $j=0$
or $j=1$ parts with respect to this choice.  Then substituting $\omega ^1=\omega
^0\eth\lambda -\lambda\eth\omega ^0+\alpha\eth\omega
^0+\beta{\overline\omega}^{0'}$ (and similarly for ${\acute\omega}^1$), we find
\begin{eqnarray}\label{hug}
 A_\scrS (Z,\acute Z )={{-\rmi}\over{4\pi G}}\oint\Bigl\{
 \bigl[\psi _1 +2(\psi _2+\sigma\dot{\overline\sigma} )\eth\lambda
 +\eth(\lambda (\psi _2+\sigma\dot{\overline\sigma}))\bigr]
 \omega ^0{\acute\omega}^0\nonumber\\
 +(\psi _2+\sigma\dot{\overline\sigma} )(\alpha{\acute\omega}^0\eth\omega ^0
 +\acute\alpha \omega ^0\eth{\acute\omega}^0)\nonumber\\
 +(\psi _2+\sigma\dot{\overline\sigma} )(\acute\beta \omega
^0{\acute{\overline\omega}}^{0'} +\beta
{\acute\omega}^0{\overline\omega}^{0'})
 \Bigr\}\,\rmd\scrS\, .
\end{eqnarray}
The middle term can be written as one proportional to $(\alpha +\acute\alpha
)\eth (\omega ^0{\acute\omega}^0)$ (but this has pure $j=1$ and so contributes
nothing to the integral in our choice of frame) and one proportional to $(\alpha
-\acute\alpha )({\acute\omega}^0\eth\omega ^0-\omega ^0\eth{\acute\omega}^0)
=(\alpha -\acute\alpha ) I_{\alpha\beta}Z^\alpha {\acute Z}^\beta$, which is
constant over the cut. Letting $\pi _{A'}$ and ${\acute\pi }_{A'}$ denote the
spinor fields $\omega ^0$ and ${\acute\omega}^0$, we may write in parallel
to~(\ref{ktwc}) and (\ref{faeq})
\begin{eqnarray}\label{huge}
A_\scrS (Z,\acute Z)=2\rmi\mu ^{A'B'}\pi _{A'}{\acute\pi}_{B'}
 +(\rmi /2) M (\alpha -\acute\alpha )\epsilon ^{R'S'}\pi
_{R'}{\acute\pi}_{S'}\nonumber\\
 +\rmi P^{AA'}(\acute\beta \pi _{A'}{\acute{\overline\pi}}_A
 +\beta{\overline\pi}_A{\acute\pi}_{A'})\, ,
\end{eqnarray}
where the three terms correspond to the three lines of~(\ref{hug}). (The
notation here is conventional, but there is one point where it can be
confusing.  This is that the symbol $\pi _{A'}$ now stands for the spinor field
$\omega ^0$ as an abstract element of $\Spin _{A'}$, and not for the spinor
field $\pi _{1'}\omicron _{A'}-\pi _{0'}\iota _{A'}$ representing components of
the local twistor, as it did previously.  From this point on, we shall only use
$\pi _{A'}$ to stand for an element of $\Spin _{A'}$, and we shall write out the
spinor field $\pi _{1'}\omicron _{A'}-\pi _{0'}\iota _{A'}$ when we need it.)

The formula~(\ref{spin}) for the spin requires certain restrictions on the
twistors. First, we require $Z$ and $\acute Z$ to be real.  We have $\Phi
(Z)=(\omega ^0{\overline\pi}_0 +\omega ^1{\overline\pi}_1+\pi
_{0'}{\overline\omega}^{0'}+\pi _{1'}{\overline\omega}^{1'})\Bigr| _\gamma
=(-\rmi\omega ^1\eth
'{\overline\omega}^{0'}+\rmi{\overline\omega}^{1'}\eth\omega ^0)\Bigr| _\gamma
=\rmi (\lambda -\alpha -\overline\lambda +\overline\alpha )\eth\omega
^0\eth'{\overline\omega}^{0'}\Bigr| _\gamma$.  Thus $Z$ is real iff $\alpha
-\lambda (\gamma )$ is real.  (And similarly $\acute Z$ is real iff
$\acute\alpha -\lambda (\acute\gamma )$ is real.)

The special-relativistic formula for the spin also requires $A_\scrS (Z)
=A_\scrS (Z,Z)\allowbreak =0$ and $A_\scrS(\acute Z)=A_\scrS (\acute Z ,\acute
Z)=0$.  The force of these conditions can be deduced from~(\ref{huge}), however,
by letting the twistors coincide.  We find $\mu ^{A'B'}\pi _{A'}\pi _{B'} +\beta
P^{AA'}{\overline\pi}_A\pi _{A'}=0$, which we regard as an equation for $\beta$,
and similarly for $\acute\beta$.

With these formulas in hand, it is straightforward to show that
\begin{eqnarray}\label{spine}
\Re A_\scrS (Z,\acute Z)/I(Z,\acute Z) &=& (2P^{AA'}{\overline\pi}_A\pi
_{A'}
 P^{BB'}{\acute{\overline\pi}}_B{\acute\pi}_{B'})^{-1}\cdot\nonumber\\
&&\Bigl[ (S_{\rm v}^{AA'}+MP^{AA'}\Im\lambda (\gamma ))
 {\overline\pi}_A\pi _{A'} P^{BB'}{\acute{\overline\pi}}_B
{\acute\pi}_{B'} \nonumber\\
&&-(S_{\rm v}^{AA'}+MP^{AA'}\Im\lambda (\acute\gamma ))
 {\acute{\overline\pi}}_A{\acute\pi}_{A'} P^{BB'}{{\overline\pi}}_B
{\pi}_{B'}\Bigr]\, ,
\end{eqnarray}
where the freedom in $\alpha$, $\acute\alpha$ has canceled out and
\begin{equation}
 S_{\rm v}^{AA'}=2\Re\rmi\mu ^{A'B'}P^A{}_{B'}
\end{equation}
is the usual formula for the Pauli--Luba\'nski spin-vector. Remembering that in
special relativity this is $M$ times the spin, we may in our case define the
spin by
\begin{equation}
 {\rm spin}(\pi _{A'})= M^{-1}S_{\rm v}^{AA'}{\overline\pi}_A\pi
_{A'}+P^{AA'}{\overline\pi}_A\pi _{A'}\Im\lambda (\gamma )\, ,
\end{equation}
where $\gamma$ is the generator of $\scrif$ corresponding to $\pi _{A'}$.

The vectorial part $S_{\rm v}^a$ is orthogonal to the time axis here.  It is
natural therefore to replace the null vector ${\overline\pi}^A\pi ^{A'}$ by its
spatial projection.  The term $\lambda (\gamma )$ depends only on this
projection as well.  Thus with a slight abuse of notation, the spin in the unit
$\hat{\bf r}$ direction is
\begin{equation}\label{spink}
{\rm spin}(\hat{\bf r} )=
 M^{-1}S_{\rm v}^a{\hat{\bf r}}_a +M\Im\lambda (\hat{\bf r})\, .
\end{equation}

The simplicity of this formula, and the clean connection of spin and
magnetic shear it gives, have already been noted. 
That this formula should exist at all --- that the quantity within
square brackets in~(\ref{spine}) should allow a separation of the
terms in $Z$ and ${\acute Z}$ --- is also remarkable.

\subsection{Mass-moments and center-of-mass}\label{elsec}

If $P_a$ and $\mu ^{A'B'}$ are the energy--momentum and angular
momentum of a special-relativistic system, and $P_a$ is timelike
(which we will always assume), then we may write
\begin{equation}
 \mu ^{A'B'}=\left( k^{AA'}-\rmi M^{-2}S^{AA'}\right) P_A{}^{B'}\, ,
\end{equation}
where $M$ is the mass, and $k^a$, $S^a$, each orthogonal to $P_a$, represent the
center of mass (in the plane $P_ax^a=0$) and the Pauli--Luba\'nski spin vector.
Thus the center of mass and spin appear as the real and imaginary parts of a
single quantity.  In special relativity, in order to identify the center of mass
of a system, one looks for points at which the vector $k^a$ vanishes.  If we
wish to do this by examining $\mu ^{A'B'}$, we must first subtract the spin
contribution. We will follow the same strategy in general relativity.

In the general-relativistic case, we may again interpret a real twistor $Z$ as a
null geodesic $\geod$ together with a tangent spinor $\pi _{A'}$, and we may
interpret $(2\rmi )^{-1}A_\scrS (Z)$ as the $\pi _{A'}$-component of the angular
momentum about $\geod$. (This $\geod$ is then not the same as the generator
$\gamma (\omega ^0)$, but $\geod$ does meet $\scrif$ at $\gamma (\omega ^0)$.)
The natural way to subtract the spin component of the angular momentum is to
form
\begin{eqnarray}\label{mmom}
(2\rmi )^{-1} A_{\scrS} (Z)
 +\rmi M^{-1}P_A{}^{A'}\pi _{A'}{\partial\over{\partial{\overline\pi}_A}}
 {\rm spin}
=\left(\mu ^{A'B'}+\rmi M^{-2}S_{\rm v}^{AA'}P_A{}^{B'}
 \right) \pi _{A'}\pi _{B'} \nonumber\\
+\beta P^{AA'}{\overline\pi}_A\pi _{A'}
+\rmi M^{-1}P_A{}^{A'}\pi _{A'}{\partial\over{\partial{\overline\pi}_A}}
 P^{BB'}{\overline\pi}_B\pi _{B'}\Im\lambda\, .\qquad
\end{eqnarray}
The operator $M^{-1}P_A{}^{A'}\pi _{A'}\partial /\partial{\overline\pi}_A$
figuring here is essentially the $\eth'$ operator acting on functions of
${\overline\pi}_A$, but this notation will be avoided here, because, while it is
`morally the same' as the $\eth'$ operator appearing elsewhere in this paper,
they differ by a factor, as will now be shown.  (Note that this operator passes
through $P^{BB'}{\overline\pi}_B\pi _{B'}$.)

We are using the symbol $\pi _{A'}$ to represent the field $\omega ^0$, thought
of as an element of $\Spin _{A'}$.  Remembering that we have chosen the
time-axis of the Bondi system aligned with the Bondi energy--momentum, we have
$P_a=Mt_a$ (where $t^a$ is the unit timelike vector determining the frame), and
thus, by basic spin-coefficient results, we have $t_A{}^{A'}\pi _{A'}$
corresponding to $\eth\omega ^0$, and so we must differentiate $\Im\lambda$ by
perturbing ${\overline\omega}^{0'}$ by $\eth\omega ^0$.  However, the potential
$\Im\lambda$ is given, not as a function of $\omega ^0$ (or
${\overline\omega}^{0'}$), but as a function of $\gamma$, the generator on which
$\omega ^0$ vanishes.  Perturbing the defining equation $\omega ^0(\gamma )=0$,
we find
\begin{equation}
 -(\delta\gamma _a )m^a\eth '\omega ^0 -(\delta\gamma _a ){\overline m}^a
  \eth\omega
^0 +\delta\omega ^0 =0\, ,
\end{equation}
where $\delta\gamma ^a$ is the infinitesimal perturbation of $\gamma$ as
$\overline{\omega}^{0'}$ is perturbed by
$\delta{\overline\omega}^{0'}=\eth\omega ^0$. However, we have $\eth '\omega
^0=0$, and so $\delta\gamma _a{\overline m}^a\eth\omega ^0=\delta\omega ^0$. 
Applying this to our perturbation $\delta{\overline\omega}^{0'}=\eth\omega ^0$,
we have $\delta\gamma _am^a=\eth\omega ^0 /\eth '{\overline\omega}^{0'}$. 
Therefore $t_A{}^{A'}\pi _{A'}\allowbreak \partial \Im\lambda /\partial
{\overline\pi}_A =-(\eth\omega ^0 /\eth'{\overline\omega}^{0'})\eth '\Im\lambda$
and
\begin{eqnarray}\label{mmmom}
(2\rmi )^{-1} A_{\scrS} (Z)
 +&\rmi M^{-1}P_A{}^{A'}\pi _{A'}{\partial\over{\partial{\overline\pi}_A}}
 {\rm spin}
=\left(\mu ^{A'B'}+\rmi M^{-2}S_{\rm v}^{AA'}P_A{}^{B'}
 \right) \pi _{A'}\pi _{B'} \nonumber\\
 &+\beta P^{AA'}{\overline\pi}_A\pi _{A'}
-\rmi
 P^{BB'}{\overline\pi}_B\pi _{B'}
 (\eth\omega ^0 /\eth'{\overline\omega}^{0'})\eth '\Im\lambda\, .
\end{eqnarray}
This quantity should be interpreted as the $\pi _{A'}$-component
of the mass-moment of the system at the geodesic $\geod$.

The center of mass is given by the vanishing of the mass-moments, and so by the
vanishing of~(\ref{mmmom}).  The vanishing of this equation gives us a formula
for $\beta$ in terms of $\pi _{A'}$, or, equivalently, in terms of the generator
of $\scrif$.  Thus for any generator of $\scrif$ we get a spinor $\pi _{A'}$,
and this spinor determines $\beta$ from the vanishing of~(\ref{mmmom}). This
choice of $\beta$, together with an admissible choice of $\alpha$ (recall we
must have $\alpha-\lambda (\gamma )$ real for $Z$ to be a real twistor)
determines a twistor from~(\ref{twfrm}).

In these formulas, the twistor $Z$ is one meeting $\scrif$ in the generator
labeled by $\pi _{A'}$ (or equivalently, determined by the vanishing of $\omega
^0$).  The freedom in choosing $\alpha$ simply corresponds to translating the
twistor up or down the generator (as follows from~(\ref{trnsn})), and is here
not very significant, as the time axis is chosen to coincide with the
Bondi--Sachs energy--momentum and this means the entire system is invariant
under such time translations.  The freedom in $\beta$ is more interesting, and
varying $\beta$ corresponds to making different choices of real twistor (that
is, of null geodesic), through the same point of $\scrif$, the complex parameter
$\beta$ representing the two real degrees of freedom in this choice.  Thus a
determination of $\beta$ by the vanishing of~(\ref{mmmom}) is precisely a
selection of a null geodesic inwards from the point of $\scrif$ in question,
this geodesic being interpretable as the one, through that point, directed
towards the center of mass of system.

The family of such geodesics determines a congruence which has an intuitive
geometric significance:  it is the congruence one would obtain by
supertranslating the Bondi system to remove the electric part of the shear (as
measured at $\scrS$). To see this, we examine the tangent spinor to the
congruence, which is (not the abstract $\pi _{A'}\in\Spin _{A'}$ but) the field
$\pi _{1'}\omicron _{A'}-\pi _{0'}\iota _{A'}$, evaluated at $\gamma$:
\begin{eqnarray}\label{kong}
\rmi (\eth\omega ^0 )\omicron_{A'}-\rmi (\eth '\omega
^1)\iota _{A'}&=&\rmi (\eth\omega ^0) \omicron _{A'}-\rmi(-\eth
'\lambda\eth\omega
^0
+\beta\eth '{\overline\omega}^{0'})\iota _{A'}\nonumber\\
&=&\rmi\eth\omega ^0\left(\omicron _{A'} +\left(\eth '\lambda
 -{{\eth '{\overline\omega}^{0'}}\over{\eth\omega ^0}}\beta\right)\iota
_{A'}\right)\, .
\end{eqnarray}
Comparing this with the vanishing of~(\ref{mmmom}), we see that $\beta$ is
chosen precisely to remove the contribution of the imaginary part of $\lambda$,
leaving only the real part, which is the angular potential for the electric part
of the shear.

This result seems very reasonable.  In the first place, it would have the effect
of removing any shear which was purely gauge --- that is, arose entirely from a
supertranslation. In the second place, that the magnetic portion of the shear
has canceled out means that the congruence meets $\scrif$ orthogonally in a
well-defined family of cuts (given by supertranslations by $\Re\lambda$).  These
may be interpreted as instants of retarded time in the rest-frame of the system,
as defined instantaneously at $\scrS$.

The results here and in the previous subsection have an overall parallelism
despite a considerable difference in detail. The spin and the center of mass are
very different sorts of quantities in general relativity, with the spin simply a
direction-dependent function, but the center of mass a null congruence. Yet the
magnetic part of the shear is essentially the $j\geq 2$ part of the spin, and
the electric part of the shear is essentially the $j\geq 2$ part of the center
of mass. Thus one may say that the shear itself codes the $j\geq 2$ part of the
angular momentum.

\subsection{Flux}

It is straightforward to compute the evolution of the angular momentum; one
simply uses the formula~(\ref{hug}) for the kinematic twistor together with the
formulas~(\ref{twfrm},\ref{trnsn},\ref{trnsnb}) for the twistors and their
evolution. (Such a direct approach is possible precisely because the angular
momentum is defined on a space $\scrT$ which is independent of the cut.)  We
shall only compute the evolution in retarded time for a fixed Bondi system, for
simplicity, although this paper's approach would allow the computation between
two arbitrary cuts.

We have
\begin{equation}
A_\scrS (Z,\acute Z)={{-\rmi}\over{4\pi G}}\oint\left\{ \psi _1\omega
^0{\acute\omega}^0 +(\psi _2 +\sigma\dot{\overline\sigma})(\omega
^0{\acute\omega}^1+{\acute\omega}^0\omega ^1)\right\}\, \rmd\scrS\, .
\end{equation}
Differentiating this, and using the relations
\begin{eqnarray}
 \dot\psi _1&=&\eth\psi _2+2\sigma\psi _3-8\pi GT_{(01)1'1'}\\
 \dot\psi _2&=&\eth\psi _3+\sigma\psi _4-4\pi GT_{111'1'}\, ,
\end{eqnarray}
we find, after a little work,
\begin{equation}\label{flux}
{{\partial}\over{\partial u}}A(Z,\acute Z)={\cal F}_{\rm matter}
+{\cal F}_{\rm Bondi}
 +{\cal F}_{\rm shift}\, ,
\end{equation}
where the three terms on the right will be discussed separately.

The first term in~(\ref{flux}),
\begin{equation}
 {\cal F}_{\rm matter}={{-\rmi}\over{4\pi G}}\oint\bigl\{
 -8\pi G T_{(01)1'1'}\omega ^0{\acute\omega}^0
-4\pi G T_{111'1'}
  [\omega ^0{\acute\omega}^1+{\acute\omega}^0\omega ^1]
 \bigr\}\,\rmd\scrS \, ,
\end{equation}
is the flux of energy--momentum and angular momentum carried away by
matter.  This term is exactly what one would expect on formal grounds.

The second term in~(\ref{flux}),
\begin{equation}\label{Bfx}
 {\cal F}_{\rm Bondi}={{-\rmi}\over{4\pi G}}\oint\bigl\{
[\psi _2+\sigma\dot{\overline\sigma} ]
 [2\omega ^0{\acute\omega}^0\eth\dot\lambda
 -\dot\lambda \eth (\omega ^0{\acute\omega}^0)]+
\dot\sigma\dot{\overline\sigma}
  [\omega ^0{\acute\omega}^1+{\acute\omega}^0\omega ^1]\bigr\}\,
\rmd\scrS \, ,
\end{equation}
is a generalization of the Bondi energy-loss term. It is $\T (\scrS )$-linear. 
Its contribution to the energy--momentum is through the term
$\dot\sigma\dot{\overline\sigma} [\omega
^0{\acute\omega}^1+{\acute\omega}^0\omega ^1]$, and this is exactly what one
expects and is second-order in the gravitational radiation.  Its contribution to
the angular momentum is through the term $[\psi _2+\sigma\dot{\overline\sigma} ]
[2\omega ^0{\acute\omega}^0\eth\dot\lambda -\dot\lambda \eth (\omega
^0{\acute\omega}^0)]$ (and energy--momentum parts can mix in as well, of course,
because of the origin-dependence).  This portion will in general have
contributions which are {\em first-order} in the gravitational radiation, since
$\dot\lambda$ is first-order.  What is necessary for such first-order
contributions is that the angular dependences of $\dot\lambda$ and the mass
aspect $[\psi _2+\sigma\dot{\overline\sigma} ]$ should combine to produce $j=0$
or $j=1$ terms.  This means that the angular dependence of the gravitational
radiation should be suitably correlated with the anisotropy of the system.

The final term in~(\ref{flux}),
\begin{eqnarray}\label{fnP}
{\cal F}_{\rm shift}&=&
{{-\rmi}\over{4\pi G}}\oint\Bigl\{
 [\psi _2+\sigma\dot{\overline\sigma} ]
 [\dot\lambda (\acute\gamma ) \omega ^0\eth{\acute\omega}^0
 +\dot\lambda (\gamma ) {\acute\omega}^0\eth\omega ^0\nonumber\\
 &&+(\eth\omega ^0\eth '\dot\lambda /\eth    
'{\overline\omega}^{0'})\Bigr| _{\acute\gamma} \omega
^0{\acute{\overline\omega}}^{0'}
 +(\eth\omega ^0\eth '\dot\lambda /\eth     
'{\overline\omega}^{0'})\Bigr| _{\gamma}{\acute\omega}
^0{{\overline\omega}}^{0'}]\Bigr\}\, \rmd\scrS\nonumber\\
\phantom{{\cal F}_{\rm shift}}
&=&(-\rmi /2)M\left(\dot\lambda (\acute\gamma )-\dot\lambda (\gamma
)\right) \epsilon ^{A'B'}{\acute\pi}_{A'}\pi _{B'} \nonumber\\
 &+&\rmi {\acute\pi}_{A'}t_A{}^{A'}{\partial\over{\partial
{\acute{\overline\pi}}_A}}\dot\lambda (\acute\gamma
)P^{AA'}{\acute{\overline\pi}}_A\pi _{A'}
 +\rmi {\pi}_{A'}t_A{}^{A'}{\partial\over{\partial    
{{\overline\pi}}_A}}\dot\lambda (\gamma
)P^{AA'}{{\overline\pi}}_A{\acute\pi}_{A'}\qquad
\end{eqnarray}
(where the simplifications in the last step make use of the results of the
preceding subsection), contains the terms reflecting the shift in linear
structure on $\T (\scrS )$ as $\scrS$ is evolved. These terms are in general
first-order in the radiation. Note that this term is {\em not}
origin-dependent:  it depends only on the $\omega ^0$, ${\acute\omega}^0$ parts
of the twistors, not on $\omega ^1$, ${\acute\omega}^1$.  Thus these `shift'
contributions to the flux only displace one fibre of $\scrT$ relative to
another; they do not alter the affine structure within each fibre which codes
the origin-dependence.

It is of some interest to compare the last formula for ${\cal F}_{\rm shift}$
with our earlier formula~(\ref{polar}) for the special-relativistic kinematic
twistor:  we had
\begin{equation}
 A(Z,\acute Z )=2\rmi \mu ^{A'B'}(x_{\rm
av})\pi
_{A'}{\acute\pi}_{B'} -(\rmi /2)P_ax_{\rm diff}^a\pi
_{B'}{\acute\pi}^{B'}\, ,
\end{equation}
where $x^a_{\rm av}=(x^a+{\acute x}^a)/2$, $x^a_{\rm diff}=x^a-{\acute x}^a$,
and $x^a$, ${\acute x}^a$ were any two points in Minkowski space on the real
twistors $Z$, $\acute Z$.  The general-relativistic formula for the flux has a
parallel structure, with $M(\dot\lambda (\gamma )-\dot\lambda (\acute\gamma
))\epsilon ^{A'B'}{\acute\pi}_{A'}\pi _{B'}$ corresponding to the term involving
$x^a_{\rm diff}$, and the other terms in~(\ref{fnP}) to an average of the
`shift' contributions to the flux of angular momentum at the two points.

\subsection{Nearly stationary systems}

It is worthwhile examining the evolution of the angular momentum in the case of
nearly stationary systems, partly because these account for a class of
substantial interest, and partly in order to clarify some of the ideas involved.

We must first make precise what we mean by a nearly stationary system.  Since we
are interested in the asymptotic behavior of the field only, what we have in
mind is a system which departs from stationarity only to first order near
$\scrif$.  (The system may have a strong time-dependence in the interior of
space--time.)

We take the Bondi system to be aligned with the (approximate) stationarity, so
that $\psi _2 =-M$ to zeroth order, where $M$ is the mass.  We assume as well
that the Bondi frame has been chosen so that the shear itself (and not just its
$u$-derivative) is a first-order quantity.  It is no loss in generality to
assume that $\lambda$ is chosen to have only $j\geq 2$ components, as before. 
And we shall assume, for simplicity, no matter fields are present near
infinity.  Then the flux ${\cal F}_{\rm matter}$ due to matter vanishes.

The Bondi flux term ${\cal F}_{\rm Bondi}$ also vanishes. First, the terms
proportional to $\dot\sigma\dot{\overline\sigma}$ and
$\sigma\dot{\overline\sigma}$ are second-order, and so will be neglected here. 
This leaves only the terms proportional to $\psi _2$. However, for the
particular spherical harmonics considered here, these must vanish, because the
only first-order terms represent products of $M$ (pure $j=0$) with $\dot\lambda$
(only $j\geq 2$) and $\omega ^0{\acute\omega}^0$ (only $j=0$, $j=1$).

We are left with only the `shift' flux term ${\cal F}_{\rm shift}$. The $\pi
_{A'}$-component of the total emitted angular momentum will be
\begin{eqnarray}
 -(2\rmi )^{-1}\Delta A(Z)&=&-(2\rmi )^{-1}\int _{u_0}^{u_1}
 {\cal F}_{\rm shift}\Bigr| _{Z=\acute Z}\, \rmd
u\nonumber\\
 &=&-P^{AA'}{\overline\pi}_A\pi _{A'}\pi
_{A'}t_A{}^{A'}{\partial\over{\partial{\overline\pi}_A}}\Delta\lambda\,
.
\end{eqnarray}
Comparing this with the discussions of the center of mass and spin in the
preceding subsections, we see that {\em if a system departs only to first order
from stationarity near $\scrif$, then the energy--momentum and the $j=1$ parts
of the angular momentum are unchanged,} but the $j\geq 2$ parts of the angular
momentum change with the shear, the electric and magnetic parts of the shear
contributing to the change in center of mass and spin, respectively. This
implies that {\em if in addition the system is stationary initially and finally,
then the change in angular momentum is a pure $j\geq 2$ change in the center of
mass, with no change in the spin,} because the magnetic component of the shear
vanishes in any stationary regime.

\section{Linearized theory and special relativity}\label{lint}

In this section, I discuss some elements of the linearized (that is,
post-Minkowskian) theory.  Of course, this linearized theory would be adequate
for many physical situations.  But there is another reason for treating it, an
important conceptual one going to the liaison between special and general
relativity, and that is the main concern here.

In fact, we seem to be confronted with a paradox.  We have been forced by
general relativity to introduce a concept of angular momentum with an entirely
new character, a non-Poincar\'e dependence on the component being measured.  On
one hand, one can understand that general relativity, whose asymptotic symmetry
group is not the Poincar\'e group, might lead to such a structure.  But (on the
other) it seems that this new, non-Poincar\'e, angular momentum is something
which never could be exchanged with non-gravitational systems, since {\em their}
angular momenta {\em are} Poincar\'e-covariant! If this really were the case,
the definition put forward here of general-relativistic angular momentum would
be very questionable.

However, this paradox is only apparent.  While it is true that {\em strictly}
special-relativistic angular momenta do not have the non-Poincar\'e character
discovered here, we shall see that {\em as soon as even linearized generic
gravitational effects are considered, consistency} requires {\em the use of the
non-Poincar\'e angular momenta.} Thus as soon as one admits any sorts of
general-relativistic corrections one is forced to non-Poincar\'e angular
momenta. We shall see that the consistent application of these ideas, far from
causing paradoxes, resolves a conundrum that had existed.

The situation might be compared to the transition from classical to quantum
mechanics.  In classical mechanics, the angular momentum is a c-number which has
a definite value at any time.  However, as soon as one treats an angular
momentum quantum-mechanically, one is forced to an entirely new sort of object,
an operator.  And as soon as one angular momentum is treated 
quantum-mechanically, one must in general (unless special simplifications apply)
treat all angular momenta which couple to it quantum-mechanically.

\subsection{Two sorts of linearization}

There are in fact two different sorts of linearization one might speak of, and
it will be important to distinguish them.  In each case, the gravitational field
is treated as a perturbation of Minkowski space. However, in one case the
twistors are simply taken to be those of Minkowski space, whereas in the other
the twistors are also perturbed to first order.  Roughly speaking, the first
case corresponds to treating gravity as a helicity-two field on Minkowski space,
whereas the second corresponds to the linearization of the full, nonlinear
theory of angular momentum.  It is the second case which is relevant here.

\subsection{First-order corrections to special relativity}

Let us consider a system with no incoming gravitational radiation. If $T_{ab}$
represents the linearized stress--energy tensor, then the retarded solution to
the linearized Einstein equation has
\begin{equation}
\psi _4=-2^{3/2} G{\overline m}^b{\overline m}^d{{\partial ^2}\over
{\partial u^2}}\int\delta\left(
l_a(ut^a-y^a)\right) T_{bd}(y)\,\rmd ^4y\, .
\end{equation}
Since $\psi _4=-\ddot{\overline\sigma}$, we expect, then,
\begin{equation}\label{linshr}
\overline\sigma =2^{3/2} G{\overline m}^b{\overline m}^d
\int\delta\left(
l_a(ut^a-y^a)\right) T_{bd}(y)\,\rmd ^4y\, .
\end{equation}
(This is not quite obvious, because of the integrations involved, but it can be
verified by direct calculations.) Thus the presence of matter generates shear at
$\scrif$, and in general this shear evolves.

As an illustration, consider perhaps the simplest case, the scattering of
particles due to contact forces.  Then (\ref{linshr}) gives the contribution of
a particle with mass $\mu$ to $\overline\sigma$ as $2^{3/2}G\mu ({\overline
m}_a{\dot\gamma}^a)^2/(l_b{\dot\gamma}^b)$, where $\dot\gamma ^a$ is the
four-velocity of the particle. Thus the shear is determined by the velocities of
the particles and their masses. If we consider the case of particles which
scatter off each other by contact, the shears before and after are in general
different.  Thus {\em even for very simple interactions, the first-order
general-relativistic correction will result in a net change in shear in a
scattering process}.  This will result in non-Poincar\'e contributions to the
angular momentum.

\subsection{Resolving a difficulty}

The treatment of angular momentum given here  resolves a conundrum having to do
with the extension of angular momentum to linearized gravity.

Suppose we try to extend the special-relativistic treatment of momentum and
angular momentum to take into account first-order gravitational effects.  We
must then deal with the fact that in general (except for very simple systems
with no interactions), the first-order corrections to a special-relativistic
system will result in the emission of gravitational waves, and also changes in
the shear.  This change in the shear breaks the Poincar\'e symmetry at $\scrif$
and introduces a first-order ambiguity in the selection of a Poincar\'e subgroup
of the BMS group and hence in the definition of angular momentum.

It is at this point that attempts to incorporate first-order
general-relativist\-ic corrections to theories of angular momentum run into
difficulties.  The ambiguity in the choice of Poincar\'e subgroup leads to all
the familiar difficulties with defining angular momentum in the full theory
(although at a linearized level).  In other words, taking into account general
relativity to first order, the angular momenta of several billiard balls before
and after a collision are not related by a Poincar\'e motion, because the
gravitational waves emitted in the collision distort the geometry of space--time
after the collision relative to that before it. (Of course, the error in
neglecting this is very small, amounting to a positional ambiguity $\sim
GM\Delta v^2/c^4\sim 10^{-42}$ cm!)

The present treatment resolves this difficulty, for the angular momentum is
defined twistorially and the shift in shear is accommodated.  We see that a
conventional mechanical system would be expected to emit small amounts of
gravitational radiation, and that radiation would give a small non-Poincar\'e
correction to the angular momentum.  It is only when we completely neglect
general-relativistic effects, and model the system special-relativistically,
that the usual Poincar\'e structure is really recovered.

\section{Number of independent constants of motion}\label{constm}

In special relativity, the angular momentum comprises six independent constants
of motion (assuming the energy--momentum is given).  The definition given here
of angular momentum in general relativity seems to comprise {\em infinitely
many} constants of motion, because it consists of apparently independent
contributions for each $j\geq 1$. Have we in fact found that somehow general
relativity, which one would expect to have {\em fewer} symmetries than special
relativity, really has infinitely many {\em more} constants of motion?

Surprisingly, there is an important sense in which the answer to this question
is Yes.  However, the situation requires some discussion.

First, we are not considering absolute constants of motion (those, in fact,
would be less interesting to us), because we wish to treat a system from which
material and gravitational radiation may be escaping. One should in fact think
of a choice of cut as partitioning the system into two portions, an interior one
(determined by data on a partial Cauchy surface meeting the cut), and the
portion of $\scrif$ to the past of the cut.  We are also especially interested
in those constants of motion which can be computed from fields on the cut, for
physically of course those represent data accessible asymptotically at a
particular retarded time.

Second, for the systems we consider, there are infinitely many constants of
motion (for general relativity is a field theory). Of all these constants of
motion, the most natural and interesting for the radiative modes are those
forming the shear at a cut.  (So there is a sphere's worth of constants of
motion, being the values of the shear at the different generators.)  These are
locally measurable at the cut, and they are physically interesting also as the
time-integrals of the wave profiles.

If we count the number of constants of motion, we have six for the $j=1$,
`conventional', angular momenta, and $2(2j+1)$ for the $j\geq 2$ degrees of
freedom in the shear.  What is new is really not the degrees of freedom in the
shear, but our {\em interpretation} of these as representing angular momenta.

That said, how is it that we have acquired infinitely many more constants of
motion than there were in special relativity? The answer is that we have taken
into account the gravitational degrees of freedom, and the coupling between
gravity in matter, and these are neglected in special relativity, and so we now
have available to us infinitely many degrees of freedom, and constants of
motion, which were not contemplated in special relativity.

Suppose, for example, we considered some system in special relativity, and
computed the integral~(\ref{linshr}) that we know, from linearized general
relativity, would lead to the shear.  In special relativity we would not call
this a constant of motion, and we would have no particular reason to be
interested in it.  However, in linearized gravity it gives us the shear, and we
interpret its change over time as giving us information about the gravitational
radiation.  It is the fact that we keep track of this gravitational radiation,
with its functional degrees of freedom, which makes the difference:  the change
in shear is accounted for as a transfer of a quantity of these constants of
motion from the internal portion of the space--time to the emitted gravitational
radiation.

In short, the count of constants of motion comes out as expected; it is rather
the unification of the concepts of shear and angular momentum which is new.

\section{Conclusions}\label{fc}

The main conclusion of this paper is that it is possible to treat the angular
momentum of an isolated gravitational system by introducing a suitable twistor
space.  This space is naturally defined in terms of the universal structure of
null infinity $\scrif$.  It lacks the full linear structure special-relativistic
twistor space has, but it does possess a fiber-bundle structure.  The base space
in question is the space of spinors at infinity, or, essentially equivalently,
the space of generators of $\scrif$.

The nonlinearities as we pass from one fibre to another in twistor space become,
in space--time terms, angular dependences for the angular momentum which are
more complex than the essentially vectorial ($j=1$) behavior in special
relativity.  The two parts of the relativistic angular momentum, the spin and
the center of mass, both have angular dependences including terms for all $j\geq
1$. The terms for $j\geq 2$ are precisely due to the shear, the electric part of
the shear determining the $j\geq 2$ parts of the center of mass and the magnetic
part of the shear determining the $j\geq 2$ parts of the spin. Additionally, the
center of mass has a direct physical interpretation as a null congruence inwards
from $\scrif$, which one thinks of as directed towards the center of the system.

Angular momentum, unlike energy--momentum, can be emitted at {\em first order}
by gravitational waves, and this applies even to the `conventional' ($j=1$)
parts.  However, such first-order emission does require a correlation between
the anisotropy of the waves and of the mass-aspect of the system, and this is
not expected for systems departing only to first order from stationarity.  For
nearly stationary systems, angular momentum can be emitted by first-order
gravitational radiation, but it has a purely $j\geq 2$ character.

The appearance of higher-$j$ terms in the angular momentum is due to the fact
that it becomes appropriate and natural to identify data for the gravitational
radiation (the shear) with some angular momenta. This radiation field has
infinitely many conserved quantities (although not absolutely conserved), given
precisely by the values of the shear, and so the theory of angular momentum is
now a theory of infinitely many such quantities.

\end{document}